\documentclass[10pt, conference, letterpaper]{IEEEtran}
\IEEEoverridecommandlockouts

\usepackage[T1]{fontenc}        % To fix the 'command \k unavailable' error related to the bibliography
\usepackage[utf8x]{inputenc}

\usepackage{amsmath,amssymb,amsthm,array,xcolor,colortbl,graphicx,multirow}
\usepackage{microtype}
\usepackage{comment}
\usepackage{balance}
\usepackage{stmaryrd}
\usepackage{import}
\usepackage{relsize}
\usepackage{scalefnt}
\usepackage{xspace}
\usepackage{balance}
\usepackage{csquotes}
\usepackage{epsfig,graphicx}  %amsthm
\usepackage{graphicx, amssymb}
\usepackage{footnote}
\usepackage[scriptsize]{subfigure}
\usepackage{tikz}
\usetikzlibrary{shapes}
\usepackage{array,xcolor,colortbl,graphicx,multirow}

\usepackage[multiple,para]{footmisc}
\allowdisplaybreaks[3]

\usepackage{cite}

\usepackage[linesnumbered,ruled,vlined]{algorithm2e}
\SetKwInput{KwData}{Input}
\SetKwInput{KwResult}{Output}
\SetKwComment{Comment}{/* }{ */}

\usepackage{wrapfig}

\usepackage{enumitem}

\usepackage{etoolbox}
\usepackage[pdfpagelabels=true,hidelinks]{hyperref}

\newtheorem{theorem}{Theorem}

\newtheorem{definition}{Definition}

\usepackage[alignedforall]{lpform}
\usepackage{epigraph}

\newcommand{\E}{\mathcal{E}}  % configurable links

\usepackage{tikz}
\usetikzlibrary{patterns, positioning,shapes.misc,matrix, arrows.meta, calc} %

\begin{document}

\title{Approximation Algorithms for Minimizing Congestion in Demand-Aware Networks
\thanks{This project is supported by the European Research Council (ERC), grant agreement No. 864228
	(AdjustNet) Horizon 2020, 2020-2025, and supported in part by NSF grants CCF-1909111 and CCF-2228995. Long Luo is supported by the National Natural Science Foundation of China (62102066), National Key Research and Development Program of China (2023YFB2904600), and Young Elite Scientists Sponsorship Program by CAST (2022QNRC001).}}

\author{\IEEEauthorblockN{Wenkai Dai\IEEEauthorrefmark{1}, Michael Dinitz\IEEEauthorrefmark{2}, Klaus-Tycho Foerster\IEEEauthorrefmark{3}, Long Luo\IEEEauthorrefmark{4} and Stefan Schmid\IEEEauthorrefmark{5}}
	\IEEEauthorblockA{\IEEEauthorrefmark{1}Faculty of Computer Science and UniVie Doctoral School Computer
		Science DoCS, University of Vienna, Austria}
	\IEEEauthorblockA{\IEEEauthorrefmark{2}Department of Computer Science, Johns Hopkins University, USA}
	\IEEEauthorblockA{\IEEEauthorrefmark{3}Department of Computer Science, TU Dortmund, Germany}
	\IEEEauthorblockA{\IEEEauthorrefmark{4}University of Electronic Science and Technology of China, P.R. China}
	\IEEEauthorblockA{\IEEEauthorrefmark{5}TU Berlin, Germany and University of Vienna, Austria}}

\maketitle

\begin{abstract}
Emerging reconfigurable optical communication technologies allow to enhance datacenter topologies with  demand-aware links optimized towards traffic patterns. This paper studies the algorithmic problem of jointly optimizing topology and routing in such demand-aware networks to minimize congestion, along two dimensions: (1)  splittable or unsplittable flows, and (2) whether routing is segregated, i.e., whether routes can or cannot combine both demand-aware and demand-oblivious (static) links.

For splittable and segregated routing, we show that the problem is generally $2$-approximable, but APX-hard even for uniform demands induced by a bipartite demand graph. For unsplittable and segregated routing, we establish upper and lower bounds of $O\left(\log m/ \log\log m \right)$ and $\Omega\left(\log m/ \log\log m \right)$, respectively, for polynomial-time approximation algorithms, where $m$ is the number of static links. We further reveal that under un-/splittable  and non-segregated routing, even for  demands of a single source (resp., destination), the problem cannot be approximated better than $\Omega\left(\frac{c_{\max}}{c_{\min}} \right)$ unless P=NP, where $c_{\max}$ (resp., $c_{\min}$) denotes the maximum (resp., minimum) capacity. It remains NP-hard for uniform capacities, but is tractable for a single commodity and uniform capacities.

Our trace-driven simulations show a significant reduction in network congestion compared to existing solutions.
\end{abstract}

\begin{table*}[htbp]
	%	\definecolor{lgray}{rgb}{0.95,0.95,0.95}
	
	\caption{Summary of our approximation upper and lower bounds on the MCRN problem (Definition~\ref{def:1}).}\label{table:overview}
	
	\begin{center}
		\renewcommand{\arraystretch}{1.1}
		\begin{tabular}{cccccc}
			
			\hline
			\hline
			\textbf{ {Approximation Bounds \&} }& \textbf{Splittable}&\textbf{Segregated} &  \textbf{Restrictions} &  \textbf{Restrictions} &\textbf{Results}\\
			%	\cline{3-6}
			\textbf{Time Complexity} & 	\textbf{ Flow} &  \textbf{Routing}& \textbf{On Demands}& \textbf{On Capacities} &\textbf{References}\\
			\hline
			
			\rowcolor{green!15}  $2$-approximation &  yes &  yes & &  & Thm.~\ref{thm:SS_approximation}\\
			%\hline
			APX-complete &  yes &  yes &  uniform and bipartite demands & &Thm.~\ref{thm: APX-hard}\\
			%	\hline
			\rowcolor{green!15}	Tractable &  yes &  yes &  single source (resp., dest.) & &Thm.~\ref{thm: tractable}\\
			
		  $O\left(\log m/ \log\log m \right)$-approximation &   no &  yes &  &  &Thm.~\ref{thm:approximation_US}\\
			%\hline
			%\hline
			\rowcolor{green!15}	Lower Bound: $\Omega\left(\log m/ \log\log m \right) $ &  no &  both  &  &  &Thm.~\ref{lem:1}\\

		Lower Bound: $\Omega\left(c_{\max}/ c_{\min} \right) $ &  both &  no  &   single source (resp., dest.)& & Thm.~\ref{thm:3}\\
			
		\rowcolor{green!15} 	NP-hard &  both &  no  &   single source (resp., dest.)& uniform & Thm.~\ref{thm: NP-hard_uniform_capacities}\\
		
		Tractable &  both &  no  &   single commodity & uniform & Thm.~\ref{thm: tractabl_nonsegregated}\\
			
			\hline

		\end{tabular}
	\end{center}
		\vspace{-1mm}
	%	\vspace{-9mm}
\end{table*}

\section{Introduction}
The popularity of data-centric applications related to e.g., business, entertainment, or artificial intelligence, let to an  explosive growth of communication traffic, especially inside datacenters.
Over the last years, great efforts have hence been put into the design of novel and more efficient datacenter network designs.
A particularly intriguing architecture is based on emerging optical communication technologies, allowing to optimize the network topology towards the traffic demand.  Such demand-aware networks are attractive as they allow to leverage the spatial and temporal structure of workloads.  More specifically, emerging demand-aware networks, whose topologies are typically hybrid, in that a static (and demand-oblivious) network is enhanced with reconfigurable (and demand-aware) links, introduce unprecedented flexibility in adapting the network topology towards the current traffic demands. In such hybrid networks, the reconfigurable links are usually enabled by optical circuit switches~\cite{8000037,cthrough,helios}, and particularly, each optical circuit switch provides reconfigurable links by establishing connections between pairs of its ports, i.e., a \emph{matching}.

Extensive past works studied  the question of how to jointly optimize topology and routing of such reconfigurable (hybrid) networks~\cite{NANCEHALL2021100621} for different networking performance metrics, e.g., latency~\cite{projector}, throughput~\cite{Longhop,stable-matching-algorithm-agile-reconfigurable-data-center-interconnect,sigmetrics22cerberus,sigmetrics23duo,sigmetrics23mars,infocom23matching}, routing length~\cite{DBLP:conf/networking/FenzF0V19,DBLP:conf/ancs/FoersterGS18,complexity,10.1145/3351452.3351464}, flow times~\cite{DBLP:conf/infocom/DinitzM20,spaa21rdcn} etc. Interestingly,  \emph{min-congestion}, a most central performance metric in traditional networks, is still not well-understood in reconfigurable networks. Avin et al.~\cite{DBLP:conf/infocom/Avin0019} and Pacut et al.~\cite{DBLP:journals/pe/PacutDLFS21} study optimal \emph{bounded-degree} topology designs, however focusing on a static optimization model in purely demand-aware network, to minimize both  the route length and the congestion. Dai et al.~\cite{cs6435} consider a hybrid model like in our paper, showing that the problem is already NP-hard for \emph{splittable} (resp., \emph{unsplittable}) and \emph{segregated} (resp., \emph{non-segregated}) routing models when the static network is a tree of height at least two, but tractable for static networks of star topologies. Zheng et al.~\cite{DBLP:conf/icpp/ZhengZGC19} introduced a greedy\nobreakdash-based heuristic algorithm for our \emph{segregated} model but on specific topologies of datacenters. However, not much more is known w.r.t.\ corresponding approximation bounds, which motivates our study.

In this paper, we are interested in the algorithmic problem underlying such hybrid demand-aware reconfigurable  network architectures. In particular, we study the question of how to jointly optimize the topology and the routing in demand-aware  networks, with the goal of \emph{minimizing congestion}. We study two different routing models commonly applied in reconfigurable networks, namely whether routing is \emph{segregated} or not, i.e., whether or not flows either have to use exclusively either the static network or the reconfigurable connections. We also consider both splittable and unsplittable flows.
\subsection{Our Contributions}
We initiate the study of approximation algorithms for minimizing congestion in hybrid demand-aware networks (for a given matrix of demands).  Our results include an overview of approximation results and complexity characterizations in general settings (outlined in Table~\ref{table:overview}), and also  a fine-grained algorithmic analysis for restricted cases.

\paragraph{Segregated Routing} We provide a mixed-integer programming formulation for segregated and un-/splittable flow models whose LP relaxation can be solved efficiently.  For splittable flows, we present a $2$\nobreakdash-approximation algorithm by a novel deterministic rounding approach, and also prove the APX-hardness even if its demands are \emph{uniform} and the \emph{graph induced by demands} is bipartite. However, we also show that the problem  becomes tractable for demands with a single source (resp., destination). For unsplittable flows,  we show that the min-congestion reconfigurable network problem cannot be approximated better than the min-congestion multi-commodity unsplittable flow problem (MCMF)~\cite{Vazirani:2001:AA:500776},  but any $\rho$-approximation algorithm based on rounding techniques for the MCMF problem can be utilized to give a $2\rho$-approximation for the reconfigurable network problem.
This implies an approximability of $\Theta\left(\log m / \log\log m \right)$ for segregated and unsplittable routing, where $m$ is the number of static links.

\paragraph{Non-Segregated Routing} Under  the splittable (resp., unsplittable) flow model,  even for demands of a single source (resp., destination),  the problem cannot be approximated better than $\Omega\left(c_{\max}/c_{\min} \right)$ unless P=NP, where $c_{\max}$ (resp., $c_{\min}$) denotes the maximum (resp., minimum) capacity on all links, and it  still remains NP-hard for  \emph{uniform capacities}, i.e., $c: \vec{E}\cup\vec{\E}\mapsto \{a\}$ for $a\in \mathbb{R}_{> 0}$. However, the problem with uniform capacities becomes efficiently solvable for demands of a single commodity under un-/splittable flow.

Our trace-driven simulations  show that our  algorithms significantly improve on state of the art methods.
\subsection{Organization}

We introduce our formal model and preliminaries in~\S\ref{sec:model}.
Our algorithmic and hardness results for the segregated model
are presented in \S\ref{sec:segregated}, followed by a study of non-segregated routing  in \S\ref{sec:non-segregated}.
We then investigate the performance of our algorithms with trace-driven simulations in $\S$\ref{sec: eveluations}. Lastly, we discuss related work in \S\ref{sec:related_work} and conclude in \S\ref{sec:conclusion}.
\section{Model and  Preliminaries}\label{sec:model} % and Preliminary Definitions

We first introduce our network model, demands and routing policies, and then formalize the min-congestion demand-aware network design problem in~\S\ref{subsec:loadp}. We then provide preliminaries for proving hardness or approximation in~\S\ref{subsec:preliminaries_approximationhardness}.

\smallskip
\noindent\textbf{Network Model.}
Let $N=(V,E,\E,c)$ be a \emph{reconfigurable (hybrid)} network \cite{solstice,eclipse} connecting  $n$ nodes $V=\{v_1,\dots,v_n\}$ (e.g., top-of-the-rack switches), using static links $E$ (usually electrically packet-switched),
where  $\left(V,E \right) $ is called the \emph{static network} of $N$.
The network $N$ also contains a set of  reconfigurable (usually optical) links  $\E$, s.t., $(V, \E)$ constitutes a \emph{complete graph} on $V$.
The graph $\left( V,E\cup\E\right) $ is a \emph{bidirected}%
\footnote{Symmetrical connectivity is the standard industry assumption for static cabling, however for reconfigurable links as well. Outside highly experimental hardware, e.g.~\cite{projector}, off-the-shelf products use full-duplex connections~\cite{calient,polatis} and this model assumption is hence prevalent, even in Free-Space Optics~\cite{firefly} proposals.}
\emph{(multi)-graph} such that two directions of each bidirected
link $\{v_i,v_j\}\in E$ (resp. $\{v_i,v_j\}\in \E$), where $v_i,v_j\in V$, work as two \emph{(anti-parallel) directed links}  $(v_i,v_j)$ and $(v_j,v_i)$ respectively.
We use the symbol $\vec{E}$ (resp. $\vec{\E}$) to denote the set of corresponding directed links of $E$ (resp. $\E$).
Moreover, a function $c: \vec{E}\cup\vec{\E}\mapsto \mathbb{R}_{\ge 0}$ defines \emph{capacities} for both directions of each bidirected link in $E\cup\E$, where the \emph{maximum} (resp., \emph{minimum}) \emph{capacity} is denoted by $c_{\max}$ (resp., $c_{\min}$). We denote \emph{uniform capacities} if $c: \vec{E}\cup\vec{\E}\mapsto \{a\}$, where $a\in \mathbb{R}_{\ge 0}$, e.g., $a=1$.
The reconfigurable network $N$ can only implement a subset of reconfigurable links $\E$, which must be a matching $M\subseteq \E$, to provisionally enhance the static network $(V,E)$. The  enhanced graph $N\left( M\right) =\left(V, E\cup M,c \right) $  determines the actual topology of the communicating network, where $N\left( M\right)$ is called a \emph{reconfigured network} and the matching $M$ is called a \emph{reconfiguration} of $N$.

\smallskip
\noindent\textbf{Traffic Demands.}
The reconfigured network should serve a certain
communication pattern, represented as a $|V| \times |V|$
communication matrix $D:=\left( d_{i,j}\right)_{|V| \times |V|} $
(\emph{demands}) with non-negative real-valued entries.
An entry $d_{i,j}\in \mathbb{R}_{\ge 0}$ represents the traffic load (frequency) or a demand
from the node $v_i$ to the node $v_j$.  With a slight abuse of notation, let $D(v_i,v_j)$ also denote a demand from $v_i$ to $v_j$ hereafter. For each demand $D(v_i,v_j)>0$, the pair $\left( v_i, v_j\right) $ is called a commodity with the source $v_i$ and the destination $v_j$. The matrix of demands $D$ is called \emph{single commodity} if it contains only one commodity, otherwise multi-commodity. A multi-commodity matrix $D$ is called single source (resp., destination) if all commodities share the same source (resp., destination).  Demands $D$ are called \emph{uniform} if all non-zero entries $d_{i,j}>0$ have the same value. The graph $G_D$ \emph{induced by} demands $D$ is defined as a simple graph $G_D=\left( V, E_D\right)$, where $E_D=\left\lbrace \left\lbrace v_i,v_j\right\rbrace : d_{i,j}+d_{j,i}>0 \right\rbrace.$

\smallskip
\noindent\textbf{Routing Models.} For  networking, \emph{unsplittable} routing requires that all flows of a demand $d_{i,j}\in D$  must be sent along a single (directed) path, while \emph{splittable} routing does not restrict the number of paths used for the traffic of each demand; For a reconfigured network, segregated routing  requires  flows of each demand $d_{i,j}\in D$ being transmitted on either the static network or the reconfigurable link  between two endpoints of this demand, 
but non-segregated routing admits reconfigurable links used as shortcuts for flows along static links~\cite{helios,cthrough}.

Hence,  there are \emph{four different routing models}: \emph{Unsplittable \& Segregated (US)}, \emph{Unsplittable \& Non-segregated (UN)},  \emph{Splittable \& Segregated (SS)}, and \emph{Splittable \& Non-segregated~(SN)}.

\subsection{Min-Congestion Reconfigurable Network Problem}\label{subsec:loadp}

\setlength{\epigraphwidth}{0.39\textwidth}
\epigraph{``\textit{As minimizing the maximum congestion level in all links is a desirable feature of DCNs~\textnormal{\cite{augmenting,DBLP:conf/infocom/HanHLX15}}, the objective of our work is to minimize the maximum link load}''}{Yang et al.~\cite{sigload} (ACM SIGMETRICS 2020)}

\smallskip
\noindent\textbf{Congestion.}
Given a reconfigured network $N(M)$ and demands $D$, let  $f:\vec{E}\cup \vec{M}\mapsto \mathbb{R}_{\ge 0}$ be \emph{a flow} serving \emph{all} demands $D$ in $N(M)$, s.t.\ the size of the net flow from  $v_i\in V$ to $v_j\in V$ equals to $d_{i,j}$ for each demand entry  $d_{i,j}\in D$, under a routing model $\tau\in \left\lbrace \text{US}, \text{UN}, \text{SS}, \text{SN}\right\rbrace$. 
For  each $d_{i,j}\in D$, let $f_{i,j}: \vec{E}\cup \vec{M} \mapsto \mathbb{R}_{\ge 0}$ denote the sub-flow of $f$ caused by the demand $d_{ij}$. Thus, the flow $f$ can be further defined as $f=\left\lbrace f_{i,j}: d_{i,j}\in D \right\rbrace $,  where $\forall e\in \vec{E}\cup \vec{M}: f(e)=\sum_{d_{i,j}\in D} f_{i,j}\left(e \right) $ and $f\left( e\right) $ can exceed $c\left( e\right) $.
We  consider \emph{loads of a flow $f$ } on both static and reconfigurable links, i.e., $\ell: \vec{E}\cup \vec{M} \mapsto \mathbb{R}_{\ge0}$.
\emph{The load of each directed link} $e\in \vec{E}\cup \vec{M}$ induced by a flow $f$ is defined as  $\ell\left(e \right):= \frac{ f(e) }{c\left( e\right) }  $, and the \emph{maximum load} of $f$ is defined as ${\ell_\textnormal{max}(f)}:=\max\left\lbrace \ell\left( e\right): e\in  \vec{E}\cup \vec{M} \right\rbrace $. Given a routing model $\tau\in \left\lbrace \text{US}, \text{UN}, \text{SS}, \text{SN}\right\rbrace$, the \emph{congestion}  in a reconfigured network $N(M)$ to serve $D$  is defined as 
\begin{equation*}
	\lambda :=\min \left\lbrace \ell_{\max}\left(f\right) :  \textnormal{a flow}\; f \; \textnormal{serving}\; D\; \textnormal{in}\;
	 N\left(M\right)\; \textnormal{under}\; \tau. \right\rbrace 
\end{equation*}
\begin{definition}[Min-Congestion Reconfigurable Network Problem (MCRN)]
	Given  a reconfigurable network $N=\left( V,E, \E, c\right) $, a routing model $\tau\in \left\lbrace \text{US}, \text{UN}, \text{SS}, \text{SN}\right\rbrace$, and a demand matrix $D$, find a  matching (reconfiguration) $M\subseteq\E$, s.t.,  the congestion $\lambda$ to  serve $D$ in the network $N\left( M\right) $ under the routing model $\tau$ is minimized. \label{def:1}
\end{definition}
\subsection{Preliminaries of Approximation Upper and Lower Bounds}\label{subsec:preliminaries_approximationhardness}
A problem has an \emph{approximation upper bound} $\alpha$ if there exists a polynomial-time $\alpha$\nobreakdash-approximation algorithm to solve it, while a problem has  an \emph{approximation lower bound} $\alpha'$ if no polynomial-time algorithm can approximate it better than $\alpha'$ unless P=NP. An approximation factor preserving reduction~\cite{Vazirani:2001:AA:500776} can be constructed to reveal the approximability between two problems.

We next introduce the classic min-congestion multi-commodity flow problem, which will be used later.

\begin{definition}[Min-Congestion Multi-Commodity Flow Problem]
	We are given a demand matrix $D$, a static (directed) network $N=(V,E)$ with the capacity function $c:\vec{E}\mapsto\mathbb{R}_{\ge 0}$,  and a routing model (splittable/unsplittable).  Our goal is to find a flow: $$f=\left\lbrace f_{ij}: \vec{E}\mapsto \mathbb{R}_{\ge 0}\; |\; d_{ij}\in D\right\rbrace ,$$  serving  $D$ under the given (splittable/unsplittable) routing model, s.t.,  the maximum load $\max_{e\in \vec{E}}{\frac{f\left( e\right)}{c\left( e\right)}}$ can be minimized, where  $f(e)=\sum_{d_{ij}\in D} f_{ij}\left(e \right)$.  \label{def:min-conge-multi-commodity}
\end{definition}
\section{Segregated Routing Model}\label{sec:segregated}
In this section, we start to study the MCRN problem under  segregated routing. We first introduce LP-based approximation algorithms for both splittable and unsplittable flow models in \S\ref{subsec: approx_algo_segregated},  and then show that the MCRN problem of splittable flow is tractable for demands of a single source (resp., destination) in \S\ref{subsec:tractable_segregated}. Finally, we discuss  approximation lower bounds for the problem under both splittable and unsplittable flow models in $\S$\ref{subsec:hardness_segregated}.
\subsection{Approximation Algorithms}\label{subsec: approx_algo_segregated}
Our approximation algorithms are based on solving a linear programming relaxation of the MCRN problem under segregated routing and obtaining a feasible solution by deterministic rounding. Thus, we will first present the ILP formulation of the problem under segregated routing, which is same for both splittable and unsplittable flow.

\subsubsection{ILP Formulation}\label{subsubsec:ILP Formulation}
To simplify the ILP formulation, we  denote each node $v_i\in V$  simply by $i\in V$, e.g., a demand $d_{i,j}\in D$ has source $i\in V$ and sink $j\in V$.
For the segregated routing model, we note that each reconfigurable link $\{i,j\}\in M$ indicates that its demands $d_{i,j}$ and $d_{j,i}$ must be solely sent on $\{i,j\}\in \E$, and other demands, whose two endpoints are not connected by a reconfigurable link included in $M$, will only transfer on the static network $\left(V,E \right) $.
We can write the min-congestion reconfigurable network problem with $\tau \in \{\text{SS}, \text{US}\}$  as the following mixed-integer linear program~(MILP).
\begin{lpformulation}
	\hspace{20pt}	\lpobj{min}{\lambda}
	\lpeq[lp:demands]{\sum_{\substack{P\in \mathcal{P}_{i,j}}} f_P \ge \left( 1-z_{i,j}\right)\cdot d_{i,j} }{i,j\in V}
	\lpeq[lp:load]{\sum_{\substack{P\in \mathcal{P}: e\in P}} f_P \leq  \lambda \cdot c\left(e \right)  }{e\in \vec{E}}
	\lpeq[lp:load2]{z_{i,j} \cdot d_{i,j}\leq  \lambda \cdot c\left( (i,j) \right)  }{ (i,j)\in \vec{\E}}
	\lpeq[lP:bound]{0\leq \sum_{j\in V} z_{i,j} \leq 1}{i\in V}
	\lpeq[lP:expo_constraints]{f_P\ge 0}{P \in \mathcal{P}}
	\lpeq[lp:bidirection]{ z_{i,j}=z_{j,i}}{i,j\in V}
	\lpeq{ z_{i,j}\in \{0,1\}}{i,j\in V}
\end{lpformulation}

\noindent To interpret this MILP formulation, we give the following~notes:

\begin{itemize}
	\item If a bidirected reconfigurable link $\{i,j\}\in \E$ is included in $M$, then both directions $(i,j)\in \vec{\E}$ and $(j,i)\in \vec{\E}$ are implemented in $\vec{M}$ simultaneously. Let  each decision variable $z_{i,j}\in \{0,1\}$ indicate whether to select the directed reconfigurable link $(i,j)\in \vec{\E}$ into $\vec{M}$ for implementation.  Thus, the constraint~\eqref{lp:bidirection} ensures the simultaneous implementation of both directions for each bidirected reconfigurable link $\{i,j\}\in \E$.
	\item The variable $\lambda \in \mathbb{R}^+$ indicates the maximum load for all directed links in the reconfigured network $\left( V, \vec{E}\cup \vec{M}\right) $. Our goal is to minimize $\lambda$ by setting these decision variables $z_{i,j}$.
	\item For every two nodes $i,j\in V$, let $\mathcal{P}_{i,j}$ denote the set of all simple directed paths from $i$ to $j$ in the static network $\left( V,\vec{E}\right) $. For each (directed) static link $e\in \vec{E}$, let $c(e)$ denote its capacity. For each demand $d_{i,j}$,  the variable $f_P$ shows the flow size along a directed path $P\in \mathcal{P}_{i,j}$. Finally, let $\mathcal{P}$ be the collection of all directed paths $\mathcal{P}_{i,j}$ for every two $i,j\in V$, i.e., $\mathcal{P}=\bigcup_{i,j\in V} \mathcal{P}_{i,j}$.
	\item The constraints~\eqref{lp:demands} ensure that  each demand $d_{i,j}\in D$ being sent on the static network $(V,\vec{E})$ if $z_{i,j}=0$, otherwise $d_{i,j}$ being sent on a directed reconfigurable link $(i,j)\in \vec{M}$.
	\item The constraints of~\eqref{lp:load} and~\eqref{lp:load2} bound the maximum load on directed links in $\vec{E}\cup\vec{M}$  by $\lambda$.
	\item For the above MILP formulation, we can relax it into an LP formation in an oblivious way, by changing the integrality constraint  $\forall i,j\in V: z_{i,j}\in \{0,1\}$ to $\forall i,j\in V: 0 \leq z_{i,j}\leq 1$, which results in \emph{an integrality gap}  $\ge 2$.
	\item To ensure that all indicator variables $z_{i,j}$ lie in the matching polytope after relaxing $z_{i,j}\in \{0,1\}$ to  $0 \leq z_{i,j}\leq 1$,  blossom inequalities defined by~\eqref{lp:blossom} are usually required. However,  blossom inequalities are unnecessary for our LP  since our rounding procedures can always guarantee that the generated feasible solution is a matching. 
	\begin{lpformulation}
		\lpeq[lp:blossom]{\textit{Blossom Inequalities:} \lpnewline \sum_{i,j\in U}z_{i,j} \leq \frac{\left|U \right|-1}{2}}{U\subseteq V: \left|U \right| \,\text{is odd.}}
	\end{lpformulation}
\end{itemize}
\paragraph{Solving LP Relaxation.} After relaxing the above MILP, the corresponding LP relaxation can be solved efficiently, although it contains an exponential number of variables $\{f_P: P\in \mathcal{P}\}$ and constraints~\eqref{lP:expo_constraints}. There are different ways to do it, but we introduce an intuitive way, often employed in approximating the min-congestion multi-commodity flow problem~\cite{williamson_shmoys_2011,Vazirani:2001:AA:500776}. The original LP formulation can be transferred to an equivalent LP with \emph{a compact formulation}, where the flow for each demand $d_{i,j}\in D$ is presented on each  link, i.e., $\left\lbrace f^{i,j}_e\in \mathbb{R}_{\ge 0}: e\in \vec{E}\cup \vec{\E} \right\rbrace $, instead of along  paths in $\mathcal{P}_{i,j}$, and also satisfies flow conservation for all nodes. This compact LP clearly has a polynomial number of variables and constraints, which can be efficiently solved, while a solution of the compact LP can be transferred to a solution of its original LP with the same load on each link.

\paragraph{Deterministic Rounding.} 
Let $Z_{opt} = \left\lbrace z_{i,j}\right\rbrace_{i,j \in V}$ be an optimal solution to the LP relaxation of the above MILP.
Let $\lambda_{\text{opt}}$ denote the optimal congestion for the fractional optimal solution  $Z_{\text{opt}}$. %For each indicator variable $ z_{i,j}\in Z_{\text{opt}}$, let $\hat{z}_{i,j}\in \{0,1\}$ denote the result of rounding up $z_{i,j}$. 
Let $\lambda_{\text{min}}$ denote the minimized congestion  of the original ILP, which has $\lambda_{\text{opt}}\leq \lambda_{\text{min}}$. Now, we introduce our idea to obtain a feasible integer solution $\hat{Z} = \left\lbrace \hat z_{ij}\right\rbrace_{i,j \in V}$   based on rounding each fractional variable $z_{i,j}$ in $Z_{\text{opt}}$.

We first consider the following rounding-up function:

\begin{equation*}
	\hat{z}_{i,j} =\begin{cases}
		1 & \text{ if }  z_{i,j}> 1/2;\\
		
		0 & \text{ if }z_{i,j}\leq  1/2.
	\end{cases} \label{eq:round-up}
\end{equation*}

It is easy to see that $\hat{z}_{i,j}= 	\hat{z}_{j,i}$ for all $i,j \in V$, since $z_{i,j}=z_{j,i}$ in $Z_{\text{opt}}$.
For each $i,j\in V$, we round up its flows $f_P$ for all  $P\in \mathcal{P}_{i,j}$ by the following way:

\begin{equation*}
	\hat{f}_{P} =\begin{cases}
		f_P/\left( 1-z_{i,j}\right), & \text{ if }  \hat{z}_{i,j} =0;\\
		
		0 & \text{ if }  \hat{z}_{i,j} =1.
	\end{cases}
\end{equation*}

\subsubsection{Approximation Results}
Next, we will show several approximation results based on the above rounding-up method.
\begin{theorem}
	When $\tau=\text{SS}$, the  min-congestion reconfigurable network problem has a polynomial-time $2$\nobreakdash-approximation algorithm. \label{thm:SS_approximation}
\end{theorem}
\begin{IEEEproof}
	Due to the degree bound of one, without violating~\eqref{lP:bound}, each node $i\in V$ can have at most one indicator variable $z_{i,j}\in Z_{\text{opt}}$ that has $z_{i,j}>1/2$, where $(i,j)\in \vec{\E}$. Thus, by rounding up the variable $z_{i,j}>1/2$ to one, each node $i\in V$ can have at most one directed reconfigurable link $(i,j)\in \vec{M}$, which implies that \eqref{lP:bound} is still satisfied by $\hat{Z}$. For each $i,j\in V$, if $\hat{z}_{i,j}=1$ then the constraint~\eqref{lp:demands} is clearly satisfied as $0=0$; otherwise, by summing flows $\hat{f}_P$ for all $P\in \mathcal{P}_{ij}$, we can have the inequality~\eqref{eq:flows}, still satisfying~\eqref{lp:demands}.
	
	\begin{align}
		\sum_{\substack{P\in \mathcal{P}_{i,j}}} \hat{f}_P = \sum_{\substack{P\in \mathcal{P}_{i,j}}}\dfrac{f_P}{\left( 1-z_{i,j}\right) }  &=  \dfrac{1}{\left( 1-z_{i,j}\right) } \sum_{\substack{P\in \mathcal{P}_{i,j}}} f_P \nonumber\\ &\ge   \dfrac{\left(1-z_{i,j} \right)d_{i,j}}{\left( 1-z_{i,j}\right) } =d_{i,j} \label{eq:flows}
	\end{align}
	
	For  each directed static link $e\in \vec{E}$, we consider the constraint~\eqref{lp:load}. Note that $\hat{f}_P>0$ holds, where $P\in \mathcal{P}_{i,j}$ and $i,j\in V$, only if $z_{i,j}\leq 1/2$, which implies $\hat{f}_P=f_P/(1-z_{i,j})\leq 2f_P$. Thus, we have the following inequation.
	
	\begin{equation}
		\sum_{\substack{P\in \mathcal{P}:e\in P}}\hat{f}_P \leq 	\sum_{\substack{P\in \mathcal{P}:e\in P} }2\cdot f_P \leq 2\cdot \lambda_{\text{opt}} \cdot c(e)\label{eq:ratio1}
	\end{equation}
	
	Regarding the constraint~\eqref{lp:load2}, for  each $(i,j)\in \vec{\E}$, it implies   $$\hat{z}_{i,j}\cdot d_{i,j}\leq 2\cdot z_{i,j}\cdot d_{i,j}\leq 2\cdot \lambda_{\text{opt}}\cdot c\left( (i,j)\right).$$
	Now, it is safe to conclude that $\hat{Z}$ is a feasible integer solution, and $f:=\left\lbrace \hat{f}_P: P\in \mathcal{P}_{i,j} \text{ and } i,j \in V\right\rbrace $ defines a feasible flow serving all demands $D$. As $\lambda_{\text{opt}}\leq \lambda_{\text{min}}$, it implies that our deterministic rounding-up method can achieve  a $2$-approximation ratio  within $O\left( n^2\right) $. The exact runtime of the algorithm depends on the specific LP formulation and LP solver, but is polynomial.
\end{IEEEproof}
We will now extend the above method to obtain an approximation result for the MCRN problem of  $\tau=\text{US}$.
\begin{theorem}\label{thm:approximation_US}
	If  the  min-congestion multi-commodity unsplittable flow problem has a $\rho$\nobreakdash-approximation algorithm based on rounding techniques on its LP solution, e.g., $\rho=O\left(\log m/ \log\log m \right) $~\cite{10.1007/s11235-016-0190-2}, then the MCRN problem with $\tau=\text{US}$ can be approximated by $2\rho$. 
\end{theorem}
\begin{IEEEproof}
	First, we note that the ILP formulation in \S\ref{subsubsec:ILP Formulation} also works for $\tau=\text{US}$. Let $\lambda_{\text{opt}}$ be the optimal value of the above relaxed LP for $\tau=\text{SS}$, which is obviously a lower bound for the MRCP problem with  $\tau=\text{US}$.
	
	For the MCRN problem of $\tau=\text{US}$, we first solve it by assuming $\tau=\text{SS}$ as Theorem~\ref{thm:SS_approximation} to obtain  a matching $M$. For each $\{i,j\}\in M$, we set $d_{i,j}=0$ and $d_{j,i}=0$ in $D$ to obtain a new set of demands $D'$. By replacing $D$ by $D'$ into the ILP formulation in \S\ref{subsubsec:ILP Formulation}, we can obtain a new relaxed LP formulation $LP\left( D'\right) $, which has an optimal  congestion $\lambda$. Clearly, it has $ \lambda\leq 2\cdot \lambda_{\text{opt}}$ by Theorem~\ref{thm:SS_approximation}. Given $D'$, we can solve the min-congestion multi-commodity unsplittable flow problem on a static network $\left(V,E \right) $ to obtain a $\rho$-approximation congestion $\lambda'$ by applying rounding techniques on the splittable (optimal) flow of  $LP\left( D'\right) $, which further implies $\lambda'\leq  \rho\cdot \lambda \leq 2\rho \cdot\lambda_{\text{opt}}$. Thus,  $\lambda'$ is a $2\rho$-approximation result for   $\tau=\text{US}$.
\end{IEEEproof}
\subsection{Polynomial-Time Solvable Cases}\label{subsec:tractable_segregated}
After obtaining approximation results for general cases of segregated model, we will show a restricted case, where demands contain a single source (resp., dest.), is efficiently solvable in Theorem~\ref{thm: tractable}. The proof of  Theorem~\ref{thm: tractable} is easy as only one reconfigurable link can be used here under segregated model, which is deferred due to limited space.
\begin{theorem}\label{thm: tractable}
	When $\tau=\text{SS}$, if the demands $D$ have a single source (resp., dest.),  the  MCRN problem is polynomial-time solvable.
	%\vspace{-2mm}
\end{theorem}
\subsection{Approximation Lower Bounds for Segregated Routing Model}\label{subsec:hardness_segregated}
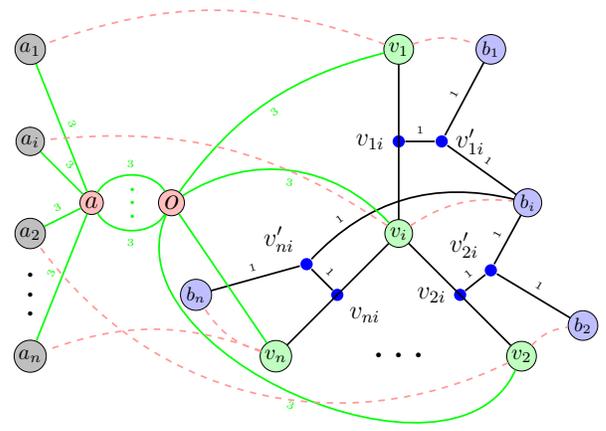
\begin{figure}[!ht]
	\begin{center}
		\resizebox{0.9\columnwidth}{!}{	
			\begin{tikzpicture}[shorten >=0.2pt]
				\usetikzlibrary{calc}
				
				\tikzstyle{vertex}=[circle,draw=black,fill=black!25,minimum size=10pt,inner sep=1pt, outer sep=0pt]
				\tikzstyle{vertex2}=[circle,draw=black,fill=blue!25,minimum size=10pt,inner sep=1pt, outer sep=0pt]
				\tikzstyle{vertex3}=[circle,draw=black,fill=green!25,minimum size=10pt,inner sep=1pt, outer sep=0pt]
				\tikzstyle{vertex4}=[circle,draw=black,fill=red!25,minimum size=10pt,inner sep=1pt, outer sep=0pt]
				
				\tikzstyle{destination}=[circle,draw=black,fill=yellow!25,minimum size=10pt,inner sep=0pt, outer sep=0pt]
				\tikzstyle{edge}=[thick,black,--,]
				
				\begin{scope}[shift={(-1,0)}]
					
					\node[vertex3] (v_{1}) at (0,0) {$v_i$};
					\node[vertex3] (v_{2}) at (0,3) {$v_1$};
					\node[vertex3] (v_{3}) at (2,-2) {$v_2$};			
					\node[vertex3] (v_{n}) at (-2,-2) {$v_n$};
					%	\node[vertex2] (v_{12}) at (0,1.5) {\large$v_{12}$};					
					%	\node[vertex2] (v_{13}) at (1,-1) {\large$v_{13}$};		
					%	\node[vertex2] (v_{1n}) at (-1,-1) {\large$v_{1n}$};

					\node[vertex2] (u_{1}) at (2.1,0.5) {\small$b_i$};
					\node[vertex2] (u_{2}) at (1.5,3) {\small$b_1$};
					\node[vertex2] (u_{3}) at (3,-1.5) {\small$b_2$};			
					\node[vertex2] (u_{n}) at (-3.3,-1) {\small$b_n$};

					%		\node[vertex2] (vin) at ($(v_{1})!0.5!(v_{n})$) {\large$v_{i,n}$};

					%					\node[circle,color=red, inner sep=2pt,fill,label= below  right:\large$v_{in}$] (vin) at ($(v_{1})!0.5!(v_{n})$) {};

					%					\draw[red, thick] (vin) -- ++(-0.6,0.5) node[circle,color=red, inner sep=2pt,fill,label= above  left:\large$c_{in}$] (cin) {};

					\node[circle,color=blue, inner sep=2pt,fill,label=below right:\large$v_{ni}$] (dni) at ($(v_{1})!0.5!(v_{n})$) {};
					
					%					\draw[green!60!black, thick] (dni) -- ++(-0.6,0.5) node[draw=black, circle,color=green!60!black, inner sep=2pt,fill,label= above left:\large$u_{n,i}$] (uni) {};
					
					%					\path(uni) edge[thick, black, bend left=10] node[midway,sloped,above] {} (u_{n});
					
					\node[circle,color=blue, inner sep=2pt,fill,label= above left:\large$v'_{ni}$] (d-in) at ($(dni)+(-0.5,0.5)$) {};

					%					\draw[green!60!black, thick] (din) -- ++(-0.6,0.5) node[draw=black, circle,color=green!60!black, inner sep=2pt,fill,label= above  left:\large$u_{i,n}$] (uin) {};
					
					%					\path(uin) edge[thick, black, bend left=20] node[midway,sloped,above] {} (u_{1});
					
					%					\path(v_{n}) edge[thick, black, bend left=20] node[midway,sloped,above] {} (u_{n});

					\path(v_{n}) edge[thick, black] node[midway,sloped,above] {} (v_{1});
					
					\path(u_{n}) edge[thick, black, ] node[midway,sloped,above] {\tiny$1$} (d-in);
					
					\path(u_{1}) edge[thick, black, bend right=30] node[pos=0.8,sloped,above] {\tiny$1$} (d-in);
					
					\path(dni) edge[thick, black] node[midway,sloped,above] {\tiny$1$} (d-in);

					\node[circle,color=blue, inner sep=2pt,fill,label= left:\large$v_{1i}$] (d2i) at ($(v_{1})!0.5!(v_{2})$) {};
					%					\draw[green!60!black, thick] (d2i) -- ++(0.8,0) node[draw=black, circle,color=green!60!black, inner sep=2pt,fill,label= right:\large$u_{1,i}$] (u2i) {};
					
					%					\path(u2i) edge[thick, black] node[midway,sloped,above] {} (u_{2});	
					
					\node[circle,color=blue, inner sep=2pt,fill,label= right :\large$v'_{1i}$] (d-i2) at ($(d2i)+(0.7,0)$) {};
					
					%					\draw[green!60!black, thick] (di2) -- ++(0.8,0) node[draw=black, circle,color=green!60!black, inner sep=2pt,fill,label= right :\large$u_{i,1}$] (ui2) {};
					
					%		\path(ui2) edge[thick, black] node[midway,sloped,above] {} (u_{1});
					
					\path(u_{2}) edge[thick, black, ] node[midway,sloped,above] {\tiny$1$} (d-i2);		
					
					%%%%%%%%%%%%%%%%%%%%%%%%%%%%%%%%%%%%%%%%%%%%%%%%%%%%%%%%%%%%%%%%%%%%%%
					
					%					\node[circle,color=blue, inner sep=2pt,fill,label= left:\Large$v_{1,i}$] (v2i) at ($(v_{1})!0.7!(v_{2})$) {};
					
					\path(v_{1}) edge[thick, black] node[midway,sloped,above] {} (v_{2});
					%					
					% 					\path(v2i) edge[thick, black] node[midway,sloped,below] {} (vi2);
					%					
					%			\path(vi2) edge[thick, black] node[midway,sloped,above] {} (v_{2});
					
					\path(u_{1}) edge[thick, black,] node[midway,sloped,above] {\tiny$1$} (d-i2);		
					
					\path(d2i) edge[thick, black] node[midway,sloped,above] {\tiny$1$} (d-i2);				
					
					%	\node[vertex2] (v3i) at ($(v_{1})!0.5!(v_{3})$) {\large$v_{i,3}$};
					
					%		\node[circle,color=red, inner sep=2pt,fill,label= below left:\large$v_{i2}$] (v3i) at ($(v_{1})!0.5!(v_{3})$) {};

					%		\draw[red, thick] (v3i) -- ++(0.6,0.5) node[circle,color=red, inner sep=2pt,fill,label= right:\large$c_{i2}$] (c3i) {};

					\node[circle,color=blue, inner sep=2pt,fill,label=   left:\large$v_{2i}$] (d3i) at ($(v_{1})!0.5!(v_{3})$) {};

					\node[circle,color=blue, inner sep=2pt,fill,label= above   left:\large$v'_{2i}$] (d-i3) at ($(d3i)+(0.5,0.4)$) {};

					\path(v_{1}) edge[thick, black] node[midway,sloped,below] {} (v_{3});
					
					\path(u_{3}) edge[thick, black, ] node[midway,sloped,above] {\tiny$1$} (d-i3);		
					
					\path(u_{1}) edge[thick, black, ] node[midway,sloped,above] {\tiny$1$} (d-i3);
					
					\path(d3i) edge[thick, black, ] node[midway,sloped,above] {\tiny$1$} (d-i3);								

					\node at (0,-2) {\huge$\mathbf{\cdots}$};
					
				\end{scope}

				\begin{scope}[xshift=-11cm]
					\node[vertex] (a1) at (4,1.5) {$a_i$};
					\node[vertex] (a2) at (4,3) {$a_1$};
					\node[vertex] (a3) at (4,0) {$a_2$};			
					\node[vertex] (an) at (4,-2) {$a_n$};	
					\path (a3) -- (an) node [black, very thick, font=\Large, midway, sloped] {\huge$\mathbf{\ldots}$};

					\node[vertex4] (b) at (6.3,0.5) {\Large$o$};	
					
					\node[vertex4] (t) at (5,0.5) {\large$a$};	
					
					%		\draw[thick]  (t) -- (b);		

					\path(t) edge[thick, green, bend left =60] node[midway,sloped,above] {\tiny$3$} (b);	
					
					\path(t) edge[thick, green, bend right =60] node[midway,sloped,below] {\tiny$3$} (b);

					\path (5.65,1) -- (5.65,0) node [green, very thick, font=\Large, midway, sloped, ] {$\mathbf{\cdots}$};							

					\path(v_{1}) edge[thick, green,bend right=40] node[midway,sloped,below] {\tiny$3$}    (b);	
					
					\path(v_{2}) edge[thick, green,bend right=20]  node[midway,sloped,below] {\tiny$3$}  (b);	
					
					\path(v_{n}) edge[thick, green]    (b);	
					\path(v_{3}) edge[thick, green, bend left=90]  node[midway,sloped,below] {\tiny$3$}  (b);			
					
					\foreach \i in {1,...,3,n}
					{%\draw[black, very thick]  (a\i)--(t1)    node[midway,sloped,above]{$1$};
						%	\path(b\i) edge[thick, ]    (b);
						
						%		\path(b\i) edge[thick, red, dashed, bend right=20]    (a\i);
						%	\path(v_{\i}) edge[thick, red!20, dashed, bend right=20]    (a\i);

						\path(t) edge[thick, green, ] node[midway,sloped,above] {\tiny$3$}   (a\i);
						%	\path(a\i) edge[thick, black] node[ midway,sloped,above] { $1$} (a);

						\path(v_{\i}) edge[thick, red!40, dashed, bend left=20]    (u_{\i});

					}
					
					\path(v_{1}) edge[thick, red!40, dashed, bend right=20]    (a1);
					\path(v_{2}) edge[thick, red!40, dashed, bend right=20]    (a2);
					\path(v_{3}) edge[thick, red!40, dashed, bend left=40]    (a3);
					\path(v_{n}) edge[thick, red!40, dashed, bend right=20]    (an);

				\end{scope}

			\end{tikzpicture}
		}
		
	\end{center}
	\vspace{-6mm}
	\caption{Illustration of our gap-preserving reduction from the minimum vertex cover problem to the MCRN problem with $\tau=\text{SS}$.   Each green node ${v_1, v_2, v_i, \ldots, v_n}$ corresponds to one vertex in the given minimum vertex cover instance. The dashed lines show a subset of reconfigurable links $\E$, whose  two endpoints have non-zero demands in $D$, and solid lines indicate  static links $E$, where there are $k$ parallel paths  between $a$ and $o$ in the static network. Each bi-directed link in $\E\cup E$ has the same capacity in both directions, and we define capacities: $c:\vec{\E}\mapsto \{3\}$ and $c:\vec{E}\mapsto\{1,3\}$. The capacity value is marked on each bi-directed (static) link in the figure.}	\label{fig:apx_hard}
	\vspace{-1mm}
\end{figure}

In the following, we  show that the problem under the splittable and segregated routing is APX-complete, implying that our $2$-approximation algorithm achieves a tight bound. 

\begin{theorem}
	For $\tau=\text{SS}$, the min-congestion reconfigurable network problem (MCRN) is APX\nobreakdash-complete, even if the demands $D$ are uniform in size and the graph $G_D$ induced by  $D$ is acyclic and bipartite, e.g., a 3D matching. \label{thm: APX-hard}
\end{theorem}
\begin{IEEEproof}
	Let  $\Pi^{\text{con}}$ denote the min-congestion reconfigurable network problem (MCRN) of  $\tau=\text{SS}$, where the induced graph $G_D$ by demands $D$ is acyclic and bipartite, and $\forall d_{i,j}\in D: d_{i,j}\in \{3,0\}$\footnote{To have integer capacities, in this proof, we use a matrix of entries in $\{0,3\}$, which can be transferred to a $0/1$ matrix by scaling down each capacity  by a factor $3$.}.  We note that the minimum vertex cover problem on $3$-regular graph $\Pi$ is APX\nobreakdash-complete, which cannot be approximated better than a ratio $\rho\in (1,2)$~\cite{10.1007/3-540-62592-5_80}.
	To show APX-hardness of $\Pi^{\text{con}}$, we give a gap-preserving reduction from the minimum vertex cover problem on $3$-regular graph to reveal that $\Pi^{\text{con}}$ cannot be approximated better than a factor $\min\{\rho, 1.2\}$. 	The construction is illustrated in Fig.~\ref{fig:apx_hard}.

	Given an instance $I=\left( G_U=(U,E_U)\right) $ of the minimum vertex cover problem $\Pi$, where  $G_U$ is a $3$-regular graph with  $|U|=n$, we construct an instance  $I'$ of $\Pi^{\text{con}}$, as shown in Figure~\ref{fig:apx_hard}. Let the constructed instance  be $I'=(G,\mathcal{E}, c, D)$, where  a static network $G=(V,E)$,    a set of reconfigurable links  $\mathcal{E}$, the capacity function $c:\vec{E}\cup\vec{\E} \mapsto \{1,3\}$ (each bi-directed link has the same capacities on both directions), and  demands $D$.
	We construct  nodes $V$ as follows:
	\begin{itemize}
		\item 	For each vertex $u_i\in U$ of $I$, where $i\in \{1,\ldots, n\}$, we construct three nodes: $v_i\in V_1$, $a_i\in A$ and $b_i\in B$;
		\item For each edge $\{u_i, u_j\}\in E_U$, we construct  $2$ nodes: $v_{ij}\in V_2$ and $v'_{ij}\in V_2$;
		\item   Moreover, we have additional two nodes $\{a,o\}$.
		\item  Finally, let constructed nodes be $$V=A\cup B\cup V_1\cup V_2\cup \{a,o\} .$$
	\end{itemize}
	All static links (edges) $E$ will be constructed as follows, where their capacities are defined by $c:\vec{E} \mapsto \{1,3\}$:
	
	\begin{itemize}
		\item For  each edge $\{u_i,u_j\}\in E_U$ of $I$, where $i,j\in \{1,\ldots, n\}$ and
		$i\neq j$,  we construct the following static (bi-directed) links: $\{v_i, v_{ij}\}\in E_1$,  $\{v_{ij}, v_{j}\}\in E_1$, $\{v_{ji}, v'_{ij}\}\in E_2$,  $\{v'_{ij},b_j\}\in E_3$ and $\{v'_{ij}b_i\}\in E_3$
		\item For each  $u_i\in U$ of $I$,  there are two static bi-directed links: $\{a_i,a\}\in E_A$ and $\{v_i, o\}\in E_o$.
		\item There are $k$  paralleling  links between $a$ and $o$, i.e., $E_{a,o}=\left\lbrace \{a,o\}_i: 1\leq i\leq k\right\rbrace$.
		\item Finally, let $E=E_1\cup E_2\cup E_3\cup E_A\cup E_o\cup E_{a,o},$ and $\forall \{u,v\}\in E_1\cup E_2\cup E_3: c\left((u,v)\right)=c\left((v,u)\right)=1 $, otherwise $c\left((u,v)\right)=c\left((v,u)\right)=3$.
	\end{itemize}
	
	Moreover, we define  demands $D$ as follows:  for each  $i\in \left\lbrace 1, \ldots, n\right\rbrace $,  it has demands $D\left( v_i, b_i\right) =\delta_{v_i}$ and $D\left( v_i, a_i\right) =3$,  where $\delta_{v_i}=3$ denotes the degree of $u_i$ in $G_U$ of $I$.
	We note that the graph $G_D$ induced by demands $D$ in our construction is also a  $3$D-matching.
	
	Finally, for every $u,v\in V$, there is a reconfigurable (bi-directed) link $\{u,v\}$ with capacities: $c\left( \left( u,v\right)  \right) =c\left( \left( u,v\right)  \right)=3$. However, due to the splittable and segregated model ($\tau=\text{SS}$), only reconfigurable links $\{u,v\}$ that have $D(u,v)+D(v,u)>0$ need to be considered for reconfiguration.

	Next, we will prove:
	\begin{itemize}
		\item If $G_U$ has a vertex cover $U^*\subseteq U$ with $|U^*|\leq k$, then   the constructed instance $I'$ has  the  minimized congestion $\lambda\leq 1$;
		\item If any vertex cover $U^*\subseteq U$  in $G_U$ has $|U^*|\ge  \rho\cdot k$, where the ratio $\rho\in (1,2)$, then  the constructed instance $I'$ has the minimized congestion $\lambda \ge  \min\{\rho, 1.2\}$.
	\end{itemize}

	We first do some analysis of the constructed instance $I'$. Due to $b=1$ and $|V_1|=n$, the reconfiguration (matching) $M$ can contain at most $n$ reconfigurable links. Let $M_1\subseteq M$ denote the selected reconfigurable links between $V_1$ and $A$ and $M_2\subseteq M$ denote the selected reconfigurable links between $V_1$ and $B$, where $M_1\cap M_2=\emptyset$ and $M_1\cup M_2= M$.
	
	We first claim that the set of nodes $V^*\subseteq V_1$ contained in $M_2$ must imply  a vertex cover $U^*$ in the graph $G_{U}$, where $\forall v_i\in V^* \implies u_i\in U^*$, otherwise, the maximum load factor is~$\lambda \ge 1.2$.

	If $V^*$ implies a vertex cover $U^*\subseteq U$ in $G_U$, it is easy to note that the maximum load on the static edges in $E_1\cup E_2\cup E_3$ is at most one. If $v_i\in V^*$, then  its demand $D(v_i, b_i)=3$ is sent on the reconfigurable (directed) link $(v_i,b_i)$ with $c\left( (v_i,b_i)\right)=3$ , otherwise $D(v_i, b_i)=3$ must  go through the corresponding three paralleling paths consisting of static links: $(v_i,v_{ij}, v'_{ij},b_i)$, where each $j$ indicates an edge $\{u_i, u_j\}\in E_U$ in  $G_U$.
	
	Reversely, we assume that $V^*$ does not imply a vertex cover in the graph $G_{U}$. We know at least one edge $\{u_i, u_j\}$ is not covered in $G_U$, which also means $v_i, v_j\notin V^*$. There are at most $5$ edge-disjoint paths between $\{v_i, v_j\}$ and $\{b_i, b_j\}$ on static network $G$, which provides  total capacities of $5$. Since $D\left( v_i, b_i\right)=3 $ and  $D(v_j,b_j)=3$, then $6$ units of demands need to be sent on these $5$ edge-disjoint paths, which indicates  the maximum load $\lambda  \ge 1.2$.

	After  showing that $M_2$ must imply a vertex cover $U^*$ in $G_U$, otherwise $\lambda\ge 1.2$, then we discuss the relationship between the size  $\left|M_2 \right| $ and the maximum load $\lambda$. If $G_U$ has a minimum vertex cover $U^*$ of the size $k$, then the load on any static link in $E_o\cup E_a\cup E_{a,o}$ is at most one, which implies the congestion $\lambda\leq 1$. But if all vertex covers having size $\ge  \rho \cdot k$ in $G_U$, where $\rho\in (1,2)$, then it must have the congestion $\lambda \ge\rho$.
	
	Given an arbitrary vertex cover $U^*\subseteq U$ in $G_U$, where $|U^*|\ge \rho\cdot k$, let $V^*\subseteq V$ denote the corresponding nodes in $G$ with $|V^*|\ge \rho \cdot k$. Since each $v_i\in V^*$ must be included in the reconfigurable links $M_2$ not $M_1$, then its demand $D(v_i, a_i)=3$ must be sent through static links. Thus, there are at least $3\cdot\rho \cdot k$ units of demands that must be transferred from $o$ to $a$. Furthermore, there are $k$ paralleling static links $E_{a,o}$ between $a$ and $o$, where $c:\vec{E}_{a,o} \mapsto 3$. Clearly, the maximum load $\lambda$ on directed links in $E_{a,o}$ is at least $\rho$.

	Since the minimum vertex cover problem  on  cubic graphs  is APX-complete~\cite{10.1007/3-540-62592-5_80}, there must exist a $\rho\in (1,2)$ s.t., the minimum vertex cover problem on cubic graphs cannot be approximated better than $\rho$ unless $P=NP$. Therefore, the above reduction  implies that our problem cannot be approximated better than $\min\{1.2, \rho\}$, which further implies APX-hardness.  Since  our problem can be approximated by two, then it is in APX-complete.
\end{IEEEproof}
It remains to investigate the approximation lower bound for unsplittable and segregated routing.
In Theorem~\ref{thm:approximation_US}, we reduce the MCRN problem with $\tau=\text{US}$ into a sub-problem, which belongs to  the min-congestion multi-commodity unsplittable flow problem. Now, we can formally show the connection between these two problems in terms of approximability.

\begin{theorem}
	There is an approximation factor-preserving reduction from   the min-congestion of multi-commodity unsplittable flow problem to the  MCRN problem with $\tau=\text{US}$. Thus, the lower bound $\Omega\left(\log m/\log\log m  \right) $~\cite{https://doi.org/10.1002/net.20121} on approximating  the min-congestion of multi-commodity unsplittable flow problem is also an approximation lower bound for  the MCRN problem with $\tau=\text{US}$.\label{lem:1}
\end{theorem}
\begin{IEEEproof}
	We prove it by giving a factor-preserving reduction. Given an  instance $I=\left(V,E,c,D \right) $ of  the min-congestion of multi-commodity unsplittable flow problem (Definition~\ref{def:min-conge-multi-commodity}), we will construct an instance $I'=\left(V',E',\E^\prime, c', D' \right) $ of the MCRN problem  with $\tau=\text{US}$. W.L.O.G., we assume that the minimum value in the capacity function $c$ is one.
	
	We  first construct a copy of $I$ contained in $I'$, s.t., $V\subset V'$, $E\subset E'$, $c\subset c'$ and $D\subset D'$. In addition, for each node $v_i\in  V$, we create a node $v_{i^\prime}\in V'$ and a demand $d_{i,i'} \in D'$ with  $d_{i,i'}=\alpha$, where $\alpha>0$ is a very large number.  For each node $v_i\in  V$, we construct a reconfigurable edge $\{i,i'\}\in \E^\prime$ with capacities $c'\left((i,i') \right) =\alpha$ and $c'\left((i',i) \right) =1$, and a static link $\{i,i'\}\in E'$ having the capacity one on both directions. We finish the construction of $I'$ after setting the capacities of other unmentioned reconfigurable links in $\E^\prime$ by one.

	Since $\alpha$ is very large in the instance $I'$,  for each  node $v_i\in  V$, its reconfigurable edge $\{i,i'\}\in \E$ must be included in the matching $M$, otherwise the min-congestion in $I'$ after reconfiguration can be as large as $\alpha$.  After that, we cannot add any reconfigurable link into the matching $M$ of $I'$. Now, the instance $I'$ equals to the given instance $I$ of the multi-commodity  flow problem.  Therefore,  $I'$ cannot be approximated better than $I$. 
\end{IEEEproof}
\section{Non-Segregated Routing Model}\label{sec:non-segregated}
In this section, we study the MCRN problem under non-segregated routing, where  we are particularly interested in understanding restricted cases, such as, demands of  single source (resp., destination), and single commodity,  uniform capacities, and their combinations. 
We will first show  approximation lower bounds and  NP-hardness for uniform capacities in \S\ref{subsec:approx_hardness_non-segregated}, and then introduce some tractable restricted cases in $\S$\ref{subsec:tractable_non-segregated}.
\subsection{Approximation Lower Bounds for Non-Segregated Routing}\label{subsec:approx_hardness_non-segregated}
By Theorem~\ref{thm:3}, we show that the MCRN problem under non-segreated routing has an approximation lower bound $\Omega\left( \frac{c_{\max}}{c_{\min}}\right) $  for both splittable and unsplittable flow, which is pessimistic  since the value of $\frac{c_{\max}}{c_{\min}}$ can be arbitrarily large. We  remark that the lower bounds $\Omega\left( \frac{c_{\max}}{c_{\min}}\right) $  still holds if each link has the same capacity on both directions, which can be derived by adapting the given proof.

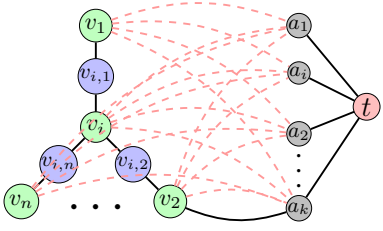
\begin{figure}[t]
	\begin{center}
		\scalebox{0.9}{	
			\begin{tikzpicture}[shorten >=0.2pt]
				\usetikzlibrary{calc}
				
				\tikzstyle{vertex}=[circle,draw=black,fill=black!25,minimum size=5pt,inner sep=0pt, outer sep=0pt]
				\tikzstyle{vertex2}=[circle,draw=black,fill=blue!25,minimum size=5pt,inner sep=0pt, outer sep=0pt]
				\tikzstyle{vertex3}=[circle,draw=black,fill=green!25,minimum size=5pt,inner sep=1pt, outer sep=0pt]
				\tikzstyle{vertex4}=[circle,draw=black,fill=red!25,minimum size=5pt,inner sep=1pt, outer sep=0pt]
				
				\tikzstyle{source}=[circle,draw=black,fill=yellow!25,minimum size=10pt,inner sep=0pt, outer sep=0pt]
				\tikzstyle{edge}=[thick,black,--,]
				
				\tikzset{
					dot/.style={circle,inner sep=2pt,fill,label={\Large #1},name=#1},
				}	
				
				\newcommand\DoubleLine[5][7pt]{%
					\path(#2)--(#3)coordinate[at start](h1)coordinate[at end](h2);
					\draw[#4, thick,]($(h1)!#1!90:(h2)$)--($(h2)!#1!-90:(h1)$)  node [black, pos=0.3, rotate=90,sloped, below=7pt] {$\mathbf{\ldots}$};
					\draw[#5, thick]($(h1)!#1!-90:(h2)$)--($(h2)!#1!90:(h1)$)  node [black,pos=0.6,sloped, above] {$=\delta\left(#2 \right) $};;
				}

				\newcommand\DoubleLinesecond[5][6pt]{%
					\path(#2)--(#3)coordinate[at start](h1)coordinate[at end](h2);
					\draw[#4, thick, shorten >=4pt, shorten <=4pt]($(h1)!#1!90:(h2)$)--($(h2)!#1!-90:(h1)$)  node [black, pos=0.3, rotate=90,sloped, below=6pt] {$\mathbf{\ldots}$};
					\draw[#5, thick, shorten >=-5pt, shorten <=-5pt]($(h1)!#1!-90:(h2)$)--($(h2)!#1!90:(h1)$)  node [black,pos=0.6,sloped, above=-2pt] {$=\delta \left( v_{n}\right) $};;
				}

				\newcommand\startingdoubleline[5][7pt]{%
					\path(#2)--(#3)coordinate[at start](h1)coordinate[at end](h2);
					\draw[#4, thick]($(h1)!#1!90:(h2)$)--($(h2)!#1!-90:(h1)$);
					\draw[#5, thick]($(h1)!#1!-90:(h2)$)--($(h2)!#1!90:(h1)$);
				}		

				\begin{scope}[shift={(-1,0)}]
					
					\node[vertex3] (v_{1}) at (0,0) {$v_i$};
					\node[vertex3] (v_{2}) at (0,1.5) {$v_1$};
					\node[vertex3] (v_{3}) at (1.1,-1.1) {$v_2$};			
					\node[vertex3] (v_{n}) at (-1.1,-1.1) {$v_n$};
					%	\node[vertex2] (v_{12}) at (0,1.5) {\large$v_{12}$};					
					%	\node[vertex2] (v_{13}) at (1,-1) {\large$v_{13}$};		
					%	\node[vertex2] (v_{1n}) at (-1,-1) {\large$v_{1n}$};	

					\node[vertex2] (vin) at ($(v_{1})!0.5!(v_{n})$) {\small$v_{i,n}$};
					
					%					\node[circle,color=blue, inner sep=2pt,fill,label=above left:\large$d_{n,i}$] (dni) at ($(v_{1})!0.7!(v_{n})$) {};
					%					
					%					\node[circle,color=blue, inner sep=2pt,fill,label=above left:\large$d_{i,n}$] (din) at ($(v_{1})!0.3!(v_{n})$) {};
					
					\path(v_{n}) edge[thick, black] node[midway,sloped,above] {} (vin);
					\path(vin) edge[thick, black] node[midway,sloped,above] {} (v_{1});
					%					
					%					\path(v_{1}) edge[thick, black] node[midway,sloped,above] {} (vin);

					\node[vertex2] (vi2) at ($(v_{1})!0.5!(v_{2})$) {\small$v_{i,1}$};

					\path(v_{1}) edge[thick, black] node[midway,sloped,above] {} (vi2);
					%					
					% 					\path(v2i) edge[thick, black] node[midway,sloped,below] {} (vi2);
					%					
					\path(vi2) edge[thick, black] node[midway,sloped,above] {} (v_{2});

					\node[vertex2] (vi3) at ($(v_{1})!0.5!(v_{3})$) {\small$v_{i,2}$};

					%					\node[circle,color=blue, inner sep=2pt,fill,label=above right:\large$d_{2,i}$] (d3i) at ($(v_{1})!0.7!(v_{3})$) {};
					%					
					%					\node[circle,color=blue, inner sep=2pt,fill,label=  above right:\large$d_{i,2}$] (di3) at ($(v_{1})!0.3!(v_{3})$) {};
					%					\node[circle,color=blue, inner sep=2pt,fill,label= below  left:\Large$v_{i,2}$] (vi3) at ($(v_{1})!0.3!(v_{3})$) {};
					\path(v_{1}) edge[thick, black] node[midway,sloped,below] {} (vi3);
					
					%					\path(v3i) edge[thick, black] node[midway,sloped,below] {} (vi3);
					
					\path(vi3) edge[thick, black] node[midway,sloped,below] {} (v_{3});

					\node at (0,-1.2) {\huge$\mathbf{\cdots}$};
					
				\end{scope}

					\begin{scope}
						\node[vertex] (a1) at (2,0.8) {\small$a_i$};
						\node[vertex] (a2) at (2,1.5) {\small$a_1$};
						\node[vertex] (a3) at (2,-0.1) {\small$a_2$};			
						\node[vertex] (an) at (2,-1.2) {\small$a_k$};	
						\path (a3) -- (an) node [black, very thick, font=\Large, midway, sloped] {$\mathbf{\ldots}$};

						\node[vertex4] (t) at (3,0.3) {\large$t$};

						\foreach \i in {1,...,3,n}
						{%\draw[black, very thick]  (a\i)--(t1)    node[midway,sloped,above]{$1$};

							\foreach \j in {1,...,3,n}{\path(v_{\i}) edge[thick, red!40, dashed, bend left=20]    (a\j);}

							\path(t) edge[thick, black, ]    (a\i);

						}
						
						\path(v_{3}) edge[thick, black, bend right=20]    (an);
						
					\end{scope}

				\end{tikzpicture}
			}
			%\vspace{-6mm}
		\end{center}
		\caption{Illustration of our gap-producing reduction from the min-vertex cover problem  to the MCRN problem in the non-segregated model. The figure shows an instance of the MCRN, where each green node ${v_1, v_2, v_i, \ldots, v_n}$ corresponds to one vertex in the min-vertex cover instance.}
		\label{fig:4}
	\end{figure}

	\begin{theorem}
		When $\tau\in \{\text{SN}, \text{UN}\}$, given demands $D$ of  a  single source (resp., destination),  the MCRN problem cannot be approximated better than   $\Omega\left( \frac{c_{\max}}{c_{\min}}\right) $  unless $P=NP$.
		%		 where $c_{\max}$ (resp., $c_{\min}$) denotes the maximum (resp., minimum) capacity defined in $c:\vec{\E}\cup \vec{E}\mapsto \mathbb{R}^+$.
		\label{thm:3}
	\end{theorem}
	\begin{IEEEproof}
		We give a proof for demands of a single destination, and  the case of a single source can be shown symmetrically.
		We give a gap-producing reduction from  a decision problem of vertex cover $\Pi$ on $3$-regular graphs to the MCRN problem $\Pi'$. For an  instance $I=(G_U=(U,E_U),k)$  of $\Pi$, where  $|U|=n$, we will construct an instance  $I'$ of  $\Pi'$, illustrated in Fig.~\ref{fig:4}.

		Let the constructed instance  be $I'=(G,\mathcal{E}, c, D)$, where  $G=(V,E)$ is the static network,    a set of reconfigurable links  $\mathcal{E}$, a capacity function $c:\vec{E}\cup\vec{\E}\mapsto \{1,\epsilon\}$, s.t., $c_{\max}=1$ and $c_{\min}=\epsilon$ ($0<\epsilon\ll 1$),  and a demand matrix $D$.
		
		The set of nodes $V$ in $I'$ are constructed as follows:
		\begin{itemize}
			\item 	For each vertex $u_i\in U$ of $I$, where $i\in \{1,\ldots, n\}$, we construct a node $v_i\in V_1$;
			\item  For each $i\in \{1,\ldots, k\}$, we have a node  $a_i\in A$ ;
			\item For each edge $\{u_i, u_j\}\in E_U$, we construct a node: $v_{i,j}\in V_2$;
			\item   Moreover, we have additional one node $t$, and let $$V=A\cup V_1\cup V_2\cup \{t\}.$$
		\end{itemize}
		Static (bidirected) links $E$ and their capacities $c:\vec{E} \mapsto \{1, \epsilon\}$ are constructed as follows:	
		\begin{itemize}
			\item For  each edge $\{u_i,u_j\}\in E_U$ of $I$, where $i,j\in \{1,\ldots, n\}$ and
			$i\neq j$,  we construct two static (bidirected) links: $\{v_i, v_{i,j}\}\in E_1$ and  $\{ v_{i,j}, v_j\}\in E_1$ with the capacities defined as:
			\begin{align*}
				c\left( \left(v_i,v_{i,j} \right) \right) =\epsilon, &\qquad\; c\left( \left(v_j,v_{i,j} \right) \right) =\epsilon,  \\
				c\left( \left(v_{i,j},v_i \right) \right) =1, &\quad\text{and } c\left( \left(v_{i,j},v_j \right) \right) =1.
			\end{align*}

			\item For each $i\in \left\lbrace 1,\ldots, k\right\rbrace $,  we have  a static (bidirected) link  $\{a_i, t\}\in E_t$, s.t., $c\left( \left( a_i,t\right) \right) =1$ and $c\left( \left(t,a_i\right) \right) =\epsilon$.
			\item Since the static network $(V,E)$ is a connected graph, we construct a static link $e^*$ connecting a node in $ V_1$ to another node in $A$, e.g., $e^*=\{v_2,a_k\}$ in Fig.~\ref{fig:4}.
			\item Finally, let $E= E_1\cup E_t\cup \{e^*\}$, where every static link has a capacity $\epsilon$ on both directions unless otherwise specified.			
		\end{itemize}
		For	any two nodes in $V$, there is a reconfigurable (bidirected) link in $\E$. For each $i\in \{1,  \ldots, n\}$ and $j\in \{i,  \ldots, k\}$,  there is one reconfigurable links $\{v_i, a_j\}\in \E$ with capacities $c\left( \left(v_i, a_j \right)  \right) =1$ and $c\left(\left( a_j,v_i \right)  \right) =\epsilon$. For other reconfigurable links, they have a capacity $\epsilon$ on both directions.

		We complete the construction by  giving demands $D$ as follows:  for each edge $\{u_i,u_j\}\in E_U$ of $I$, we have a node $v_{i,j}\in V$ with a demand $D(v_{i,j},t)=1/3$.
		
		Next, we show that, when  $\tau\in \{\text{SN}, \text{UN}\}$, if $I$  has a vertex cover of size $\leq k$, then the minimized congestion $\lambda$ of $I'$ is at most $1$; and if $I$ has a vertex cover of size $>k$, then the minimized congestion $\lambda$ of $I'$ has $\lambda\ge\frac{1}{12\epsilon}$ for $\tau= \text{SN}$ (resp., $\lambda\ge \frac{1}{3\epsilon}$ for $\tau= \text{UN}$). 
		
		Clearly, we need to find $k$ nodes in $V_1$ to connect $k$ nodes in $A$ by reconfigurable links (matching) $M$, s.t.,  each demand can be sent to $t$ without going through any directed link of the capacity $\epsilon$.

		First, if $G_U$ has a vertex cover $U'\subseteq U$ of size $k$, then for each  $u_i\in U'$, we add the corresponding  reconfigurable link $\{v_i, a_i\}\in \E$ into $M$. By this reconfiguration $M$, each node $v_{i,j}$ can  find a directed path to $t$, s.t.,  each directed link in this path has the capacity one, then $\lambda \leq 1$. 
		
		On the other hand, if $G_U$ does not  have a vertex cover  of size $\leq k$, then for any $k$ vertices  $U'\subseteq U$, there exists an edge $\left\lbrace u_i, u_j\right\rbrace \in E_U$, s.t., $u_i\notin U'$ and $u_j\notin U'$. Thus, the demand $D(v_{i,j},t)$ cannot be sent on reconfigurable links incident on either $v_i$ or $v_j$. It has to  find another node $v_f\in V_1$, which has a reconfigurable link $\left\lbrace v_f, a_l\right\rbrace $ included in a directed path $\left(v_f,a_l,t \right) $, where $a_l\in A$. 
		W.L.O.G., we assume $\left\lbrace u_f,u_i\right\rbrace \in E_U$. Then, any directed path from $v_{i,j}\in V_1$ to $v_f\in V_1$ must visit the directed link $(v_i, v_{i,f})$ that has the capacity $\epsilon$. Thus, when $\tau= \text{SN}$ (resp., $\tau= \text{UN}$), it  implies $\lambda\ge \frac{1}{12\epsilon}$ (resp., $\lambda\ge \frac{1}{3\epsilon}$), where  $c_{\max}=1$ and $c_{\max}=\epsilon$.
		
		Therefore,  the $I'$ cannot be approximated within  a ratio $\frac{1}{12\epsilon}$, i.e., $\Omega\left(\frac{c_{\max}}{c_{\min}} \right)$, unless $P=NP$.
	\end{IEEEproof}
	%
	%
	%\begin{theorem}
	%	content...
	%\end{theorem}
	After knowing the approximation lower bound $\Omega\left( \frac{c_{\max}}{c_{\min}}\right) $ in Theorem~\ref{thm:3},  an immediate question is if the problem is tractable when $c_{\max}=c_{\min}$. Next, we will negate this conjecture by showing the NP-hardness for splittable (resp., unsplittable) and segregated routing even if  it has a single source (resp., dest.) and uniform capacities. 
	\begin{theorem} \label{thm: NP-hard_uniform_capacities}
		When $\tau\in \{\text{SN}, \text{UN}\}$, the MRCN problem remains NP-hard for demands of a single  source (resp., destination) and uniform capacities.
	\end{theorem}
	\begin{IEEEproof}
		We first give a proof for  $\tau=\text{SN}$ by a many\nobreakdash-one reduction from a decision problem  $\Pi$ of the makespan scheduling on identical machines, which is NP-hard~\cite{Garey:1979:CIG:578533}. 
		
		For the makespan scheduling on identical machines~\cite{Vazirani:2001:AA:500776}, we are given a set $\mathcal{M}$ of machines and  a set $\mathcal{J}$ of jobs,  where each job $j$ has a processing time $p_{j}\in \mathbb{Z}^+$ on any machine $i$, and our goal is to schedule the jobs on the machines to minimize the \emph{makespan}, i.e., the maximum completion time.

		Given an instance $I=\left( \mathcal{J}, \mathcal{M}, \bigcup_{j\in \mathcal{J}} p_j, \beta\right) $ of $\Pi$, which decides if $I$ has a makespan bounded by $\beta$, we construct an instance $I'=\left( V,E,\E,c,D\right) $ (single destination $t$) of the MCRN problem $\Pi'$. Moreover, let $\left|\mathcal{J} \right|=n$ and $\left|\mathcal{M} \right|=m$.
		
		For each job $j\in \mathcal{J}$, we construct a source node $v_j\in V$, which has a static link $\left\lbrace v_j,t\right\rbrace \in E$ to the single destination $t\in V$ and a demand $D\left( v_j,t\right)=\beta+p_j.$ For each machine $i\in \mathcal{M}$, we construct a node $u_i\in V$ that has $n$ neighbors $U_i=\left\lbrace u^1_i, \ldots, u^n_i\right\rbrace \subseteq V$ included in the static network, and a static link $\left\lbrace u_i,t\right\rbrace \in E$. Moreover, we construct a node $t'\in V$, which has a static link $\{t,t'\}\in E$ and a demand $D\left(t',t \right) =2\beta$. Recall that, for our network model, there exists a reconfigurable link in $\E$ for any two nodes in $V$
		but only a subset of $\E$ can be selected into the matching $M\subset \E$. For the uniform capacities, we set $c:\vec{E}\cup\vec{\E}\mapsto \left\lbrace 1\right\rbrace $.
		
		Next, we need to prove that the optimal congestion $\lambda$ of $I'$ satisfies $\lambda \leq \beta$ iff the minimum makespan of $I$ is not more than $\beta$. The proof details are  deferred to the technical report due to space constraint.

		For $\tau=\text{UN}$, we can change the construction by  replacing each static link $\left\lbrace v_j,t\right\rbrace \in E$ with two static links $\left\lbrace v_j,v'_j\right\rbrace \in E$ and $\left\lbrace v'_j,t\right\rbrace \in E$ and defining new demands $D\left(v_j,t \right) =p_j$ and $D\left(v'_j,t \right) =\beta$.  Then, the above argument works for the new construction similarly.
	\end{IEEEproof}
	
	\subsection{Polynomial-Time Solvable Cases}\label{subsec:tractable_non-segregated}
	In contrast to Theorem~\ref{thm: NP-hard_uniform_capacities}, Theorem~\ref{thm: tractabl_nonsegregated} reveals that the MCRN problem with $\tau\in \{\text{SN}, \text{UN}\}$ can become tractable if demands are further restricted to be single-commodity (the proof is deferred due to space constraint).
	\begin{theorem} \label{thm: tractabl_nonsegregated}
		For   $\tau\in \{\text{SN}, \text{UN}\}$,  given a single-commodity demand and uniform capacities, the MCRN problem is polynomial-time solvable.
	\end{theorem}
%
%%%
\section{Evaluations}\label{sec: eveluations}
We complement our theoretical analysis with an empirical evaluation of the performance of our algorithms under realistic workloads. We first describe our methodology in~\S\ref{subsec: methodology}  and then discuss our results in~\S\ref{subsec: results}. We will share our implemnetation with the research community together with this paper.
\subsection{Methodology}\label{subsec: methodology}
We employ the following baselines and implemented the corresponding algorithms 
to compare with our algorithm (labelled as \emph{MC}) under \emph{segregated} and \emph{un/-splittable} models.

\smallskip
\noindent \textbf{Baselines.}
We first consider a \emph{Maximum Weight Matching} algorithm as a baseline, which aims  to maximize
the sum of flow quantities on reconfigurable links, also employed by e.g.,~\cite{cthrough,helios}.
Second, we also compare to a \emph{Greedy} approach (labelled as \emph{Greedy}), where we greedily seek a compatible reconfigurable link to reroute the maximum demands
in each iteration until the matching cannot be extended further. Similar greedy algorithms are also used by e.g., Halperin et al.~\cite{augmenting} and Zheng et al.~\cite{DBLP:conf/icpp/ZhengZGC19}. Lastly, we additionally plot the congestion on the static network without any reconfigurable link 
(label: \emph{Oblivious}) and the optimum of the LP before rounding (label: \emph{LP}) as upper and lower bounds respectively.

\smallskip
\noindent \textbf{Traffic Workloads.}
Since traffic traces in different networks and running different applications
can differ significantly~\cite{dc-nature,roy2015inside,benson2010network,sigmetrics20complexity,projector},
we collected a number of real-world
and synthetic traces to generate traffic matrices
to evaluate our algorithms.
More specifically:
\begin{itemize}
	\item \textbf{HPC traces:} We first employ four traces of exascale applications in High Performance Computing (HPC) clusters~\cite{hpcURL,sigmetrics20complexity}:  MOCFE, NeckBone, CNS, and MultiGrid.
	\item \textbf{Synthetic traces:}
	We further  consider the synthetic pFabric traces, which are frequently used as benchmarks in scientific evaluations~\cite{pfabric}.  In a nutshell, we generate demand matrices from  workloads that arrive according to a Poisson process between random sources and destinations. 
\end{itemize}
\smallskip
\noindent \textbf{Experimental Setup.}
We implement the topologies of static networks by generating random $k$-regular graphs for $k=4,8$ ($k$ denotes the number of ports connecting other ToR switches) as Jellyfish~\cite{jellyfish},
and using 
uniform capacities on both static and reconfigurable links. For unsplittable\footnote{Due to the large size of networks and demands, in practice, it is usual to restrict the maximum number of paths when computing and implementing splittable flow, as done in, e.g., Jellyfish~\cite{jellyfish}.} 
(resp., splittable) flows, we consider one shortest path (resp., three shortest paths) for each demand for HPC traces (Fig.~\ref{fig:result-hpc}) (resp., pFabric traces (Fig.~\ref{fig:result-pFabric})). We repeat each setting by running it $5$ times 
to obtain averaged results, normalizing the loads on links.

\subsection{Results and Discussion}\label{subsec: results}
\begin{figure}[t!]
\centering
\subfigure{\centering \includegraphics[width=0.84\columnwidth]{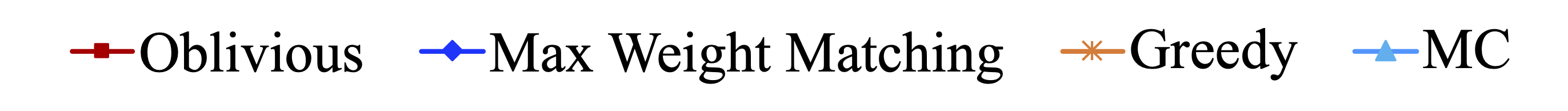}}
\hfill

\vspace{-0.4cm}
\subfigure[$k=4$]{\label{subfig:hpc-degree=4}
	\centering
	\includegraphics[width=0.47\columnwidth]{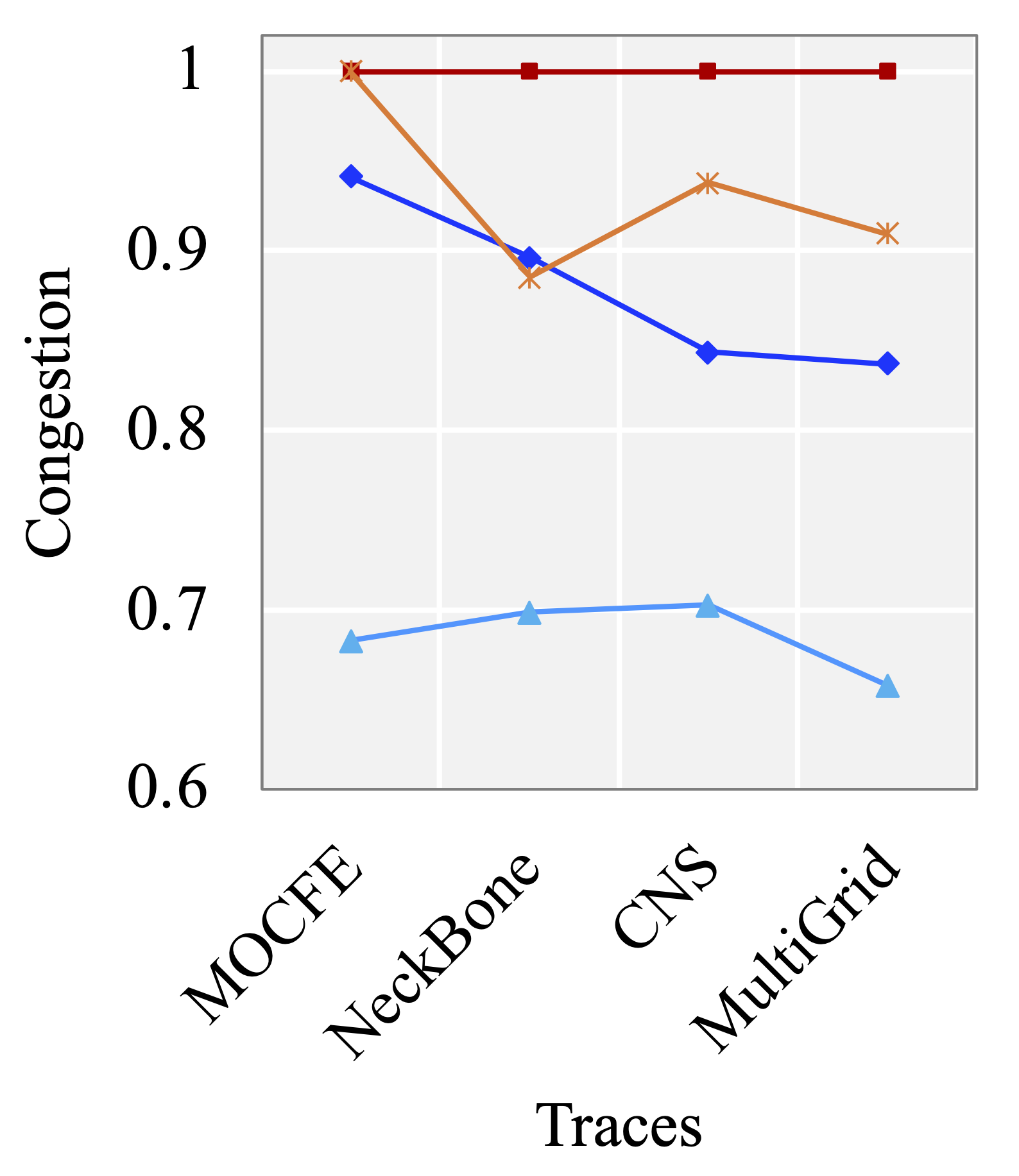}}
\subfigure[$k=8$]{\label{subfig:hpc-degree=8}
	\centering
	\includegraphics[width=0.47\columnwidth]{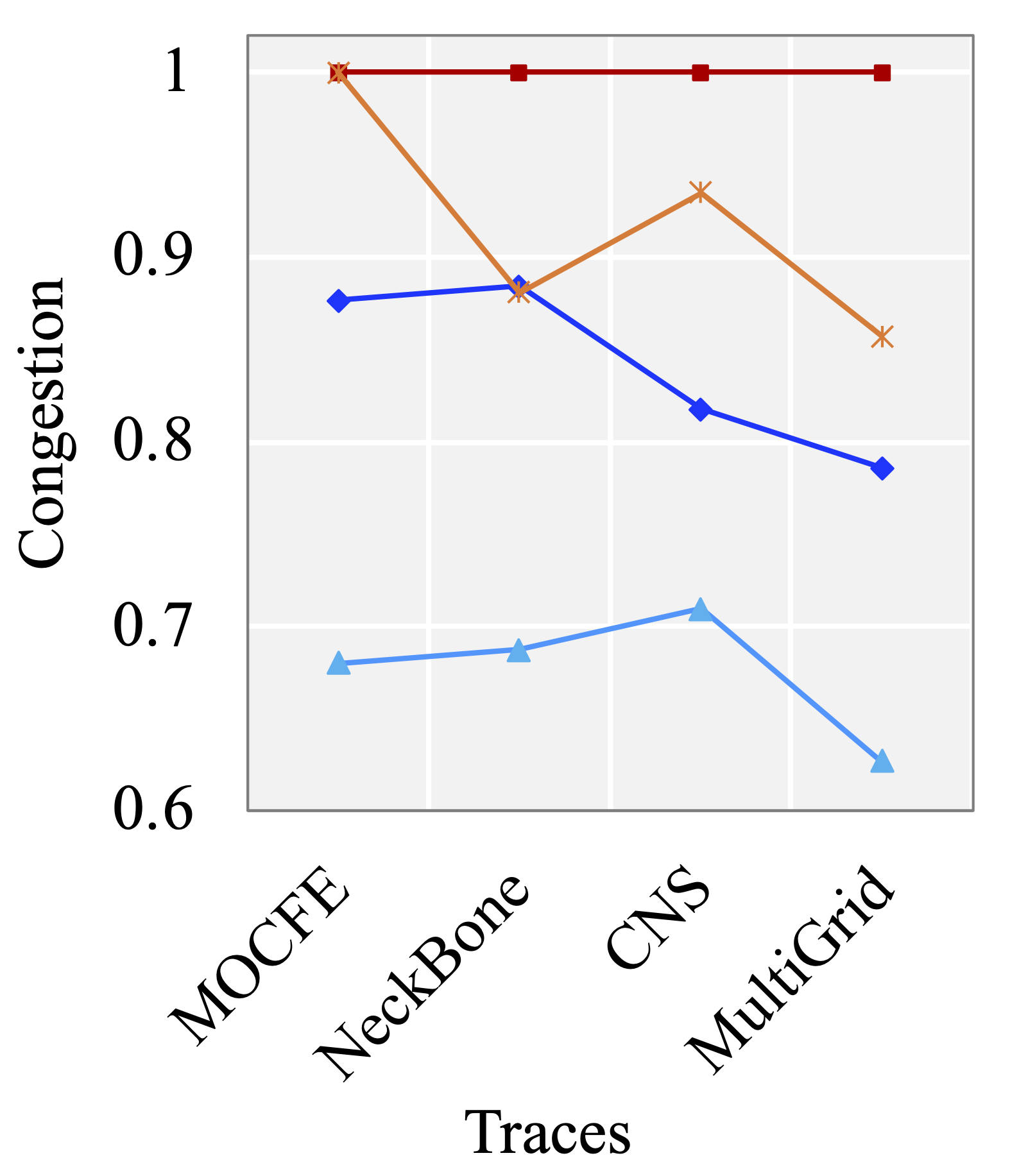}}
\caption{Algorithmic comparison of the min-congestion for four traces of exascale applications in HPC clusters, using random $k$-regular graphs for the static network.}
\label{fig:result-hpc}
\end{figure}

\begin{figure}[t]
\vspace{-0.1cm}
\centering
\includegraphics[width=0.9\columnwidth]{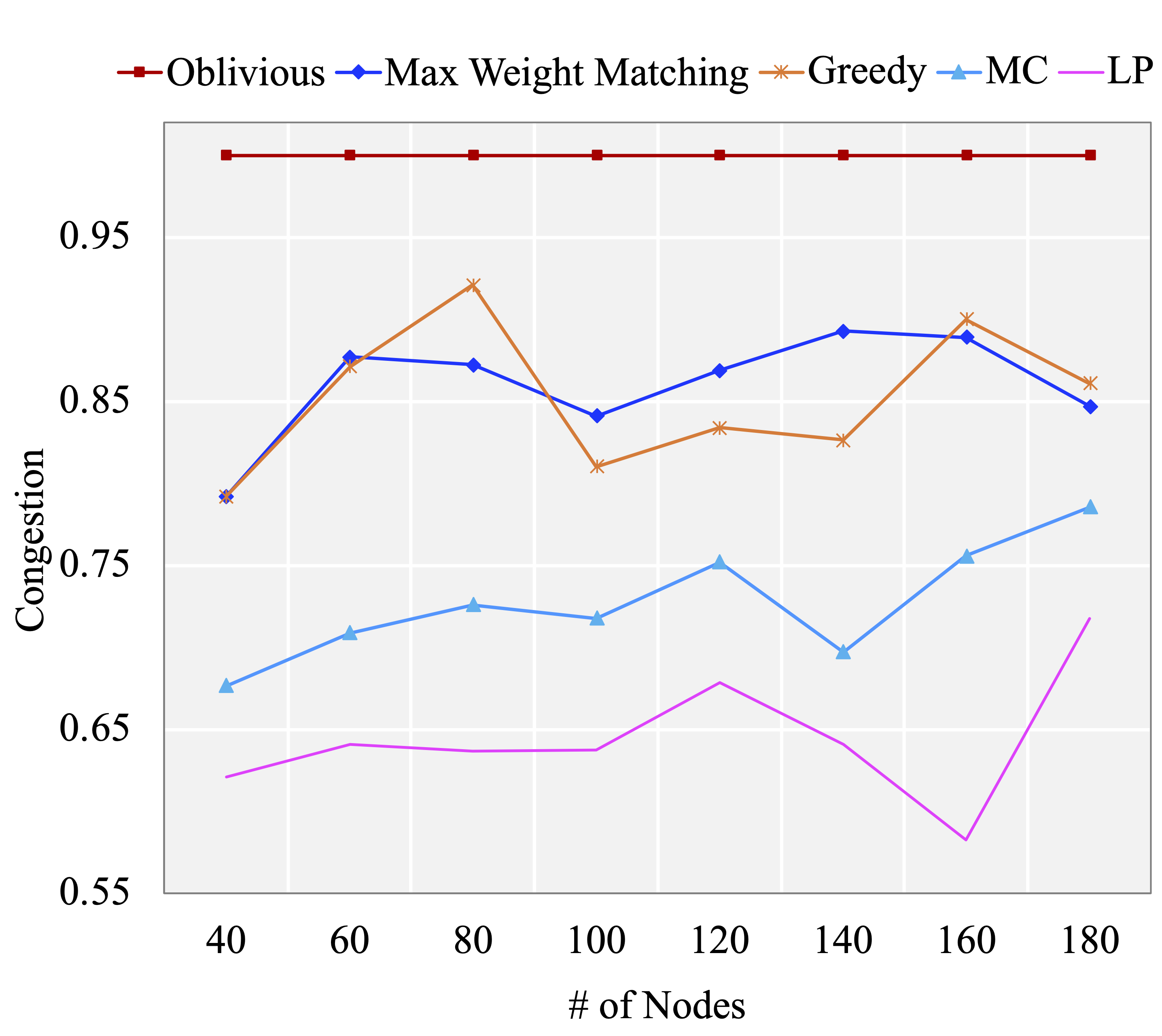}
\caption{Algorithmic comparison of the min-congestion for different synthetic traces of pFabric clusters, with random $4$-regular graphs as static networks.}
\label{fig:result-pFabric}
\vspace{-0.1cm}
\end{figure}

We summarize our evaluation results in terms of the minimized congestion for these four HPC traces in Fig.~\ref{fig:result-hpc} and for pFabric traces in Fig.~\ref{fig:result-pFabric}.

Clearly, all considered algorithms significantly improve the congestion  over the Oblivious baseline, and our algorithms (MC) typically outperform the others. We observe that our MC algorithm can provide  the stable benefits  throughout all investigated scenarios, e.g., the changing number of nodes, diverse traces and varying average degrees of static networks, while Greedy is worst in achieving a consistent performance. 

More specifically, for pFabric traces, our MC algorithm (splittable) can achieve $\approx 65\%-75\%$ of the original congestion of Oblivious baseline for $40-180$ nodes, where its performance  is best on $150$ nodes and notably decreases when nodes are increased from $140$ to $180$. The MC algorithm can at least provide a $1.3$-approximation w.r.t. the optimal value of our LP. Regarding the HPC traces, the MC algorithm (unsplittable) still dominates  Greedy and Max.\ Weight Matching, which keeps a comparatively lower congestion,  reducing the congestion of Oblivious by  $\approx 30\%-35\%$,  but the variance is slightly higher than in the  pFabric traces, matching empirical observations on the
complexity of the traces produced by these synthetic traces~\cite{sigmetrics20complexity}. Meanwhile, we can observe very similar plots for all algorithms in  static networks of $k=4$ and $k=8$.

\section{Further Related Work}
\label{sec:related_work}

Reconfigurable networks have recently received much attention both in the theory and the applied community. Reconfigurable networks generally come in two flavors: demand-oblivious networks such as RotorNet~\cite{rotornet}, Opera~\cite{opera}, Sirius~\cite{sirius}, and Mars~\cite{sigmetrics23mars} provide an unprecedented throughput by avoiding (or at least minimizing) multi-hop forwarding compared to their static counterparts~\cite{DBLP:conf/sigcomm/Al-FaresLV08,DBLP:conf/conext/ValadarskySDS16}.
Demand-aware networks such as ProjecToR~\cite{projector}, SplayNets~\cite{ton15splay}, Jupiter~\cite{poutievski2022jupiter}, Cerberus~\cite{sigmetrics22cerberus}, Eclipse~\cite{eclipse}, Duo~\cite{sigmetrics23duo}, among many other~\cite{helios,cthrough,mordia,rotornet,10.1145/3387514.3406221,NANCEHALL2021100621,spaa21rdcn,owan3,DBLP:conf/infocom/DinitzM20}, are tailored towards skewed and structured workloads and leverage spatial and temporal locality to improve throughput further~\cite{benson2010network,sigmetrics20complexity,foerster2023analyzing}.

However, many of these works revolve around other objectives or network settings, e.g.,~\cite{DBLP:conf/networking/FenzF0V19,10.1145/3351452.3351464,projector,stable-matching-algorithm-agile-reconfigurable-data-center-interconnect,Longhop,owan3,Jia2017a,spaa21rdcn,10330695,DBLP:conf/sosr/FoersterLG20,DBLP:conf/opodis/SchiffF0H22,DBLP:journals/jnca/LuoFSY22,DBLP:conf/ancs/FenzF021}, and the important aspect of congestion is still not well understood. We refer the reader to the survey by~\cite{NANCEHALL2021100621} for a more general overview.

The works by Avin et al.~\cite{DBLP:conf/infocom/Avin0019} and Pacut et al.~\cite{DBLP:journals/pe/PacutDLFS21} study bounded-degree topology designs to minimize route length and  congestion, and provide approximation lower bounds and a $6$-approximation algorithm for  min-congestion  for sparse demands. However, their models consider optimizing the topology solely consisting of reconfigurable links without a static network, which fundamentally differs from our model of hybrid networks.

Zheng et al.~\cite{DBLP:conf/icpp/ZhengZGC19} consider a problem setting similar to our segregated model and study how to enhance classic datacenter network topologies, such as Diamond, BCube and VL2, with small reconfigurable switches. They present NP-hardness results on general graphs, although these results do not transfer to specific data center topologies or trees, respectively, while they also introduce a greedy\nobreakdash-based heuristic algorithm without performance guarantees.
Yang et al.~\cite{sigload} study how to enhance the datacenter network by 60GHz wireless reconfigurable links  for min-congestion under unsplittable  and non-segregated  routing. However, their congestion definition is structurally different from our model, as they consider undirected links, and furthermore, their reconfigurable links are formed under wireless interference constraints.

Related in name and spirit, is the so-called \ensuremath{\mathsf{HYBRID}} model, introduced by Augustine et al.~\cite{DBLP:conf/soda/AugustineHKSS20}.
So far, the investigations in the \ensuremath{\mathsf{HYBRID}} model focused on, e.g., path length, diameter, and (competitive) routing~\cite{DBLP:conf/soda/AugustineHKSS20,DBLP:conf/icdcn/CastenowKS20,DBLP:conf/opodis/0001HS20,DBLP:conf/opodis/CoyC0HKSSS21,DBLP:conf/podc/KuhnS20}, and it would be interesting to develop a unifying framework of hybrid demand-aware networks and the \ensuremath{\mathsf{HYBRID}} model.

Dai et al.~\cite{cs6435} studied the same network model as we do in this paper, showing that the reconfigurable network-design for the objective of min-congestion is already NP-hard for splittable (resp., unsplittable) and segregated (resp., non-segregated) routing models when the static network is a tree of height at least two, but tractable for static networks of star topologies. However, they only provide algorithms with guarantees for very specific problem instances (i.e., star graphs) and also no approximation hardness lower bounds.
% Some preliminary results in this direction are also mentioned by Dai et al.~\cite{DBLP:conf/wdag/DaiDF022}.
%
\section{Future Work}
\label{sec:conclusion}
Our work leaves open several interesting questions for future research. In particular, it remains to provide a complete picture of tight upper and lower bounds on approximating the  non-segregated routing problems. 
\clearpage

\balance
%\vspace{-2mm}
%\addtolength{\topmargin}{0.06in}
\bibliographystyle{IEEEtran} 
\bibliography{references}

\end{document}